\documentstyle[aps,prl,multicol,epsfig]{revtex}

\begin{document}

\title{Field Theory of the Random Flux Model}

\author{Alexander Altland$^{1)}$ and B D Simons$^{2)}$} 
\address{1) Institut f\"ur Theoretische Physik, Universit\"at zu K\"oln, 
D-50937 K\"oln, Germany,\\
2) Cavendish Laboratory, Madingley Road, Cambridge CB3\ OHE, UK}

\date{\today}
\maketitle 
 
\begin{abstract} 
The long-range properties of the random flux model (lattice fermions hopping 
under the influence of maximally random {\it link} disorder) are shown to be 
described by a supersymmetric field theory of non-linear $\sigma$ model type, 
where the group ${\rm GL}(n|n)$ is the global invariant manifold. An 
extension to non-abelian generalizations of this model identifies connections 
to lattice QCD, Dirac fermions in a random gauge potential, and stochastic
non-Hermitian operators.
\end{abstract} 

\pacs{PACS numbers: 71.30.+h,72.10.-d,05.45.+b}

\begin{multicols}{2}  
Quantum disordered systems are typically realised in Hamiltonians of the 
general form $\hat H = \hat H_0 + \hat V$, where $\hat H_0$ models the 
underlying ``clean'' system, and disorder is introduced via the randomly 
distributed {\em Hermitian} operator $\hat V$. 
Sometimes, however, it is preferable to implement disorder in terms of 
{\em unitary} stochastic operators and to consider Hamiltonians of the
type
\begin{eqnarray}
\textstyle\hat{H}=-\sum\limits_{\langle ij\rangle} c^\dagger_i U_{ij} c_j,
\label{H_rf}
\end{eqnarray}
where $\langle ij\rangle$ denote neighbouring sites of a
$d$-dimensional hypercubic lattice, the $c$'s represent $N$-component
lattice fermions, and $U_{ij}$ represent $N$-dimensional unitary matrices
residing on the links of the lattice. Stochasticity is introduced by
drawing the $U$'s from a random distribution (albeit subject to the
Hermiticity requirement $U_{ij}=U_{ji}^\dagger$). Hamiltonians of the type 
(\ref{H_rf}) are commonly referred to as random flux (RF)
Models, a denotation we will also adopt for the cases $N\not=1$.

RF-models appear in a variety of different contexts: The 2d $N=1$ version 
describes the dynamics of lattice fermions subject to a 
random magnetic field or, more accurately, a random vector 
potential~\cite{numerical,analytical,Chalker,Aronov,Miller}. As well as the 
gauge theory of high T$_c$ superconducitvity~\cite{Nagaosa}, this model
has been discussed in connection with the physics of the half-filled fractional
quantum Hall phase~\cite{Halperin}, as well as the spin-split Landau 
level~\cite{Chalker,ss}. Identifying the two fermion components of the $N=2$ 
RF-model with a spin degree of freedom, (\ref{H_rf}) describes the 
propagation of lattice electrons on a spin-disordered background, a 
situation that occurs, e.g., in connection with the physics of manganese 
oxides~\cite{MuHa}. Identifying the three fermion components of the $N=3$ 
model with a color degree of freedom, (\ref{H_rf}) represents a 
prototype~\cite{fn1} of the strong coupling lattice QCD-Hamiltonian.

Superficially, (\ref{H_rf}) appears to fall into the general class of
(bond disordered) Anderson Hamiltonians. That conjecture indeed holds
true provided one stays away from the middle of the tight-binding
band, $\epsilon=0$. Upon approaching $\epsilon=0$, however, the
phenomenology of the RF-models begins to differ drastically from a
conventional disordered fermion systems. In spite of intensive
numerical and analytical 
investigation~\cite{numerical,analytical,Chalker,Aronov,Miller,ss}, 
central aspects of
these deviations are not yet fully understood.  For example, the key
question of whether or not the $2d$ RF-model possesses a band center
extended metallic phase has not yet been settled; apart from the fact
that the average density of states (DoS) diverges upon approaching
$\epsilon=0$, much of the structure of even that basic observable
remains unknown.

The purpose of this Letter is 2-fold: Firstly we provide new information 
regarding the band center behavior of the RF-model. Secondly we wish to 
discuss a diverse network of interconnections that exist between the
RF-problem and related areas of current research interest. 

Both aspects 
of that program are based on the result that the long-range behavior of 
average $n$-point Green functions, $\langle G^\pm(\epsilon+\epsilon_1)\dots 
G^\pm(\epsilon+\epsilon_n)\rangle$, of the RF-model can be obtained from a 
supersymmetric field theory defined by the action
\end{multicols}
\begin{equation}
\label{Seff}
S[T] = -\int \left[c_1{\,\rm str}(\partial T^{-1} \partial T) 
+ ic_2{\rm str\,} (\hat \epsilon (T+T^{-1})) +c_3\left({\,\rm str\,}( T^{-1}
\partial T)\right)^2\right]+S_{\rm b}[T],
\end{equation}
\begin{multicols}{2}
\noindent where $T\in GL(n|n)$ (the group of invertible supermatrices of
dimension $2n$), `str' is the standard supertrace, and the matrix
$\hat \epsilon = {\rm diag}(\epsilon_1,\dots,\epsilon_n)$. The
contribution $S_{\rm b}$\cite{fn4} represents a boundary action that
depends on the values of the fields $T$ at the corner points of the
lattice.

Eq.~(\ref{Seff}) is derived under the assumption of maximal
unitary randomness, i.e. all $U_{ij} \in {\rm U}(N)$, independently
distributed according to the Haar measure. In this case, the constants
$c_1=Na^{2-d}/8d$, $c_2=N(2d-1)^{1/2}a^{-d}/4d$, $c_3=a^{2-d}C/16d$
where $a$ represents the lattice spacing, and $C$ denotes a
geometry-dependent numerical constant $O(1)$. Below we will argue that
the structure of the field theory is actually disorder independent,
i.e. that RF-models are generally described by
(\ref{Seff})~\cite{fn2}, where the strength of the disorder manifests
itself merely in the value of the coupling constants.

Below we will outline how, starting from the `microscopic' Hamiltonian
(\ref{H_rf}), the effective description (\ref{Seff}) is derived. However,
before turning to that more technical part of the discussion, we first
address the question of what kind of information can be gained 
from the field theory. Our main goal will be to demonstrate that the 
action~(\ref{Seff}) represents a quantitative implementation of the 
network of connections displayed in Fig.~1. By exploring 
different links, we will discuss some characteristic features of the field 
theory. 
\begin{figure}
\epsfig{file=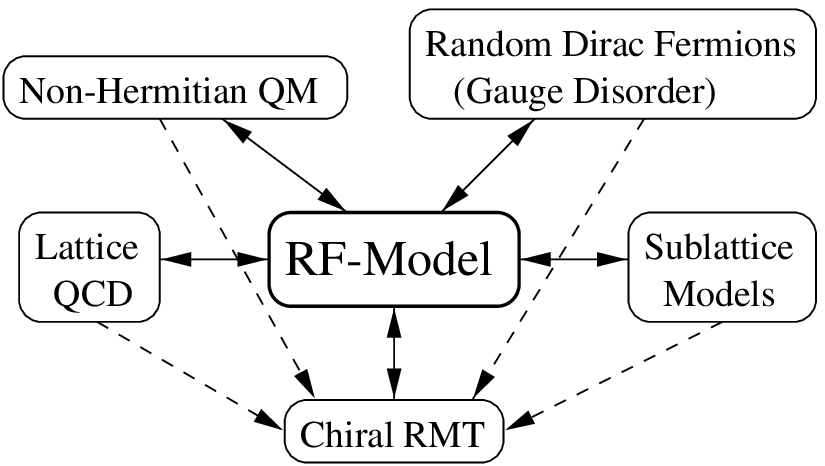}
%\vspace{0.2cm}
%\caption{\label{fig}Connection between the RF-model and various related 
%systems governed by the presence of chiral symmetries. The low energy limit
%of all models is universally described by Chiral Random Matrix
%theory.}
\end{figure}

{\small FIG. 1. Connection between the RF-model and various related 
systems governed by the presence of chiral symmetries. The low energy limit
of all models is universally described by Chiral Random Matrix
theory.}

\smallskip

{\it Chiral Random Matrix Theory (ChRMT):} As usual with field theories of
disordered systems, the low energy regime of (\ref{Seff}) (energies
$\epsilon < c_1/(c_2 L^2)$)
%, where $L$ is the linear extension of the system) 
is governed by spatially constant field configurations
$T_0={\rm const.}$,
\begin{equation}
\label{S_0}
S_0[T_0]=-{\pi \rho_0\over 2} {\rm str\,} (\hat \epsilon (T_0+T_0^{-1})) +
S_{\rm b}[T_0],
\end{equation} 
where $\rho_0$ is the bulk mean DoS of the system. Correlation functions 
computed with respect to the first contribution to the action (\ref{S_0}) 
coincide with those otherwise obtained for the chiral unitary random matrix 
ensemble ChGUE~\cite{Nagao,Hikami,Andreev,Verbaarschot}, (the ensemble of
symmetry AIII in the classification scheme of Ref.~\cite{Zirnbauera}), i.e. 
the ensemble of block off-diagonal matrices 
\begin{equation}
  \label{block}
\textstyle  \left(\begin{array}{cc}
   &A\\
A^\dagger& 
  \end{array}\right),
\end{equation}
where $A$ is complex random. In particular, the mean DoS is found to
vanish as $\epsilon \to 0$ on a scale set by the mean level spacing.
The connection to ChRMT follows readily from the fact that the RF-Hamiltonian 
possesses a chiral symmetry: Partitioned into two nested sublattices $A$ and 
$B$, the bipartite lattice Hamiltonian assumes a block off-diagonal 
form~(\ref{block}) in an $A/B$-decomposition. To the best of our knowledge, 
the ramifications of the chiral structure on the physical properties of 
RF-models was first reported in Ref.~\cite{Miller}. In passing we note that, 
in contrast to conventional disordered systems, the low energy limit of the 
RF-model is not absolutely universal: The fine structure of the DoS close to
$\epsilon=0$ depends on the `parity' of the lattice, i.e. on whether
the number of sites is even or odd~\cite{fn5}. Without going into details, 
we remark that the information about this effect is encoded in the boundary 
term $S_{\rm b}$.

{\it Non-Hermitian Operators:} To investigate problems involving
non-Hermitian stochastic operators $A\not=A^\dagger$ one commonly
introduces an operator like the one shown in (\ref{block}), i.e. an
{\it Hermitian} auxiliary operator of twice the dimension of the
original problem~\cite{Sommers88,Fyodorov}. Put differently, non-Hermitian
Hamiltonians possess an inbuilt chiral structure implying that their
low-energy universal properties coincide with those of 
manifestly chiral problems like the RF-model. That in turn means that the 
basic structure of the low energy field theory (\ref{Seff}) of the 
RF-model (a system with broken time-reversal invariance) should coincide 
with that of the (time-reversal non-invariant version of the) non-linear 
$\sigma$-model of non-Hermitian problems introduced in Ref.~\cite{Efetov}. 
To make that connection explicit, we introduce the auxiliary matrix variable
$Q=\exp(W \sigma_1/2) (\hat s \otimes \sigma_3) \exp(-W \sigma_1/2)$,
where $W=\ln T$, $\sigma_i$ are Pauli matrices operating in the block
space of (\ref{block}), and $\hat s = {\rm diag}({\rm sgn\,
  Im}\epsilon_1,\dots,{\rm sgn\, Im}\epsilon_n)$. One may check by
direct comparison that for $n=1$ (the case there considered), the
matrices $Q$ are equivalent to the degrees of freedom employed in
Ref.\cite{Efetov}. When represented in terms of $Q$'s, (\ref{Seff})
assumes a form similar to a standard~\cite{Efetovsbook} non-linear
$\sigma$-model, albeit one of novel symmetry~\cite{Efetov}.  That the
connection is not incidental, but rather extends to the more complex
variants $n>1$, follows from a) the above mentioned fact that both
non-Hermitian and RF-type problems possess a chiral structure, and b)
that only three fundamentally different $\sigma$-models with chiral
symmetry (corresponding to the cases of broken time-reversal
invariance, broken spin rotation invariance, and invariance under both
operations) exist. For a more thorough discussion of
these symmetry aspects, we refer the reader to the original
Ref.\cite{Zirnbauera}.

{\it Weakly Disordered Sublattice Models:} Leaving the random matrix regime 
and turning to the more complicated spatially extended problem, it is 
important to notice that the field theory~(\ref{Seff}) has a closely related 
precursor: Analysis of a weakly disordered sublattice model led 
Gade~\cite{Gade} to a boson replica version of the present model, i.e. a 
theory over fields $T\in {\rm GL}(nR)/{\rm U}(nR)$, where $R\to 0$ is the 
number of replicas. The action for these fields coincides with~(\ref{Seff}), 
save for the absence of the boundary term, and the important difference that, 
due to the weakness of the disorder, the coupling constant $c_1$ was 
parametrically larger than one. 

Various conclusions concerning the physical behavior of the
RF-Hamiltonian, most notably about its localization behavior, can be
inferred directly from Ref.~\cite{Gade}. There it was shown that the
conductance of the weakly disordered $2d$ model at the band center
(which is essentially determined by the coupling constant $c_1$) did
not change under one-loop perturbative renormalization. This
observation suggests that a non-localized state might exist in the
middle of the band. Since the stability of the perturbative RG relies
merely on the smallness of the parameters $1/c_1,c_3/c_1\ll 1$, its
results can be straightforwardly carried over to the $N \gg 1$
non-abelian RF model: The one-loop renormalization indicates that, at
least for $N \gg 1$, the strongly disordered RF model exhibits
metallic behavior at the band center. (It is interesting to note
that, according to the connections summarized above, the unusual
localization properties of the zero energy states of the RF-model
characterize those of {\it all} eigenstates of a stochastic
non-Hermitian operator.)

It was also predicted in Ref.~\cite{Gade} that the 
$2d$ DoS diverges upon approaching the middle of the band. In order to
understand on which energy scale that divergence sets in, and how
it will eventually be cut off deep within the random matrix regime, one
would have to superimpose perturbative RG techniques onto a
non-perturbative treatment of the low-energy regime, a task that is
well beyond the scope of the present paper.

Finally we notice that, in an RG-sense, finite energies $\epsilon_i$
represent a relevant perturbation. Renormalization of the field theory
leads to a flow away from the chiral band center limit eventually leading
to the standard unitary universality class. This result is consistent with
the analysis of Ref.~\cite{Aronov} where it was shown that continuum 
fermions (i.e. the analogue of lattice fermions close to the {\it bottom} of 
the band) subject to a weak random field map onto a unitary $\sigma$-model.

The tendency of sublattice models to exhibit band center delocalized
behavior persists even in the (quasi) $1d$ case: It was shown in 
Refs.~\cite{Miller,Brouwer} that $1d$ sublattice models with $N$ even exhibit
conventional localization behavior whilst for $N$ odd a delocalized
mode remains in the band center. This parity effect is closely related
to the odd/even phenomenon mentioned above in connection with the mean DoS, 
and indeed it is the boundary action that is responsible for the quasi $1d$ 
delocalization phenomenon within the $\sigma$-model formulation.

{\it Lattice QCD and Random Dirac Fermions: } Besides Gade's model,
the field theory (\ref{Seff}) has at least two other close relatives:
In QCD, (\ref{Seff}) has been suggested on phenomenological grounds as
relevant for the determination of the low energy spectrum of the Dirac
operator~\cite{bernard-golterman}. In that context, the base manifold
is $4+1$-dimensional whilst the fields $T\in U(n_f+1|1)$, where $n_f$
is the number of quark flavors. For a comprehensive discussion of the
QCD-analogue of (\ref{Seff}), its connection with ChRMT and its
relevance for lattice QCD-analyses, we refer the reader to Ref.~\cite{ChRMT}. 
The similarity between the theories is again a manifestation of the
universality of chiral $\sigma$-models or, more physically, the
universal consequences chiral symmetries have for the long-range
properties of random systems. (In QCD, `randomness' is represented by
gauge field fluctuations in the Yang-Mills Hamiltonian.)

Secondly, an analogue of the action (\ref{Seff}) (again without the
boundary operator $c_4$) with a Wess-Zumino-Novikov-Witten WZWN term
\begin{eqnarray*}
-c_5\int_0^1 d\zeta\epsilon^{\zeta\mu\nu}{\rm str\,}\left(T\partial_\zeta 
T^{-1}T\partial_\mu T^{-1}T\partial_\nu T^{-1}\right)
\end{eqnarray*}
was obtained from a lattice model of random Dirac
fermions~\cite{Dirac}. The connection to the RF-model follows
from the fact that the clean limit of the Dirac model can effectively
be described in terms of a model of $\pi$-flux lattice fermions. (The
Dirac-structure of the clean $\pi$-flux model is also responsible for
the occurrence of a WZMN operator.)

Having reviewed some essential features of the field theory
(\ref{Seff}), we finally outline how it is obtained from (\ref{H_rf}).
That in this Letter, the derivation is not formulated in more detail
is motivated by the observation that not only the degrees of freedom
but also the structure of the field theory is, to a large extent,
dictated by aspects of symmetry: By analogy to the situation for the
`conventional' supersymmetric $\sigma$-models~\cite{Efetovsbook},
there are only a few ${\rm GL}(n|n)$ invariant operators with $\le 2$
gradients (namely the ones appearing in (\ref{Seff}) plus the WZWN
operator~\cite{fn3}).  Thus, the `only' job that is left for a
microscopic derivation is to decide whether the operators permitted by
symmetry are actually realized in the field theory, and to fix their
coupling constants.  Here we restrict ourselves to a brief outline of
that analysis.  Details of the calculation will be presented in a
separate publication.

(i) As usual in the construction of field theories of disordered
problems, we first represent Green functions of the problem in terms
of a Gaussian integral over a field $\psi$. Choosing supersymmetry as
a way to normalize the resulting functional integrals to unity, the
first step exactly parallels the constructions reviewed in
Ref.~\cite{Efetovsbook}. (ii) Next we average over the set
$\{U_{ij}\}$. At that stage significant deviations from the standard
treatment of {\it Hermitian} disordered operators occur. A method of
exactly averaging over (extended) models involving {\it unitary}
stochasticity has been introduced in Ref.~\cite{ZirnQHE,ZirnCUE} and 
christened the `color-flavor transformation'. Following that reference we 
eliminate the disorder at the expense of introducing a pair of auxiliary fields
$\{ (Z_{ij},\tilde Z_{ij}) \}$ (which play a role analogous to the
Hubbard-Stratonovich field $Q$ commonly employed in $\sigma$-model
constructions). (iii) Integrating out the $\psi$'s we are left with the
action 
\end{multicols}
\begin{equation}
\label{Z3} 
\textstyle S[Z,\tilde Z]= 
-N\sum_{\langle i\in A, j\in B
    \rangle}{\rm str \, ln \,} 
(1-Z_{ij}\tilde Z_{ij})+
N\sum\limits_{i\in A} {\rm str \, ln \,}
\left(\hat{\epsilon} +\sum_{j\in {\sc n}_i} Z_{ij}\right)+N\sum\limits_{j\in 
B} {\rm  str \, ln \,} 
 \left(\hat{\epsilon} +\sum_{i\in {\sc n}_j} \tilde Z_{ij}\right),
\end{equation}
\begin{multicols}{2}
\noindent  
where the notation $j\in {\sc n}_i $ indicates that $j$ is summed over all 
nearest neighbours of $i$. (iv) Subjecting (\ref{Z3}) to a saddle-point 
analysis, the fields are conveniently parameterized as $(Z,\tilde Z)
\equiv (ix PT, ix T^{-1} P)$, where $x$ is a constant, and $T,P\in {\rm
GL}(n|n)$ respectively have the significance of Goldstone, massive modes of 
the theory. (v) Integrating out $P$ we find
that, in contrast to standard $\sigma$-model analyses (and in accord with
the construction of Gade's action~\cite{Gade}), a residual coupling
between massive and Goldstone modes exists; it gives rise to the
$c_3$-term in (\ref{Seff}).  (vi) The remaining, pure Goldstone action
is subjected to a gradient expansion which results in (\ref{Seff}).

Summarizing, we have derived an effective field theory for the
maximally disordered RF-model. The theory has a status
analogous to the supersymmetric non-linear $\sigma$-models of
`conventional' disordered Fermi systems, but its behavior is substantially
different, a fact that is readily traced back to the
presence of a chiral symmetry. It was shown that the formalism provides
a platform from which interconnections to a variety of other recently
investigated chiral problems can be conveniently analysed.

It is a pleasure to acknowledge valuable discussions with M. Janssen, A.
Tsvelik, and J. J. M. Verbaarschot. We thank M. R. Zirnbauer for plenty of
useful suggestions made at all stages of this work, and in particular
for pointing out the significance of the operator $O^{(1)}$.

\end{multicols}


\begin{thebibliography}{99}
\bibitem{fn1}
To obtain the strong coupling QCD-Hamiltonian, Eq.~(\ref{H_rf}) must be 
 upgraded to accommodate lattice Dirac fermions. While this additional 
 structure can be implemented within the present analysis~\protect~\cite{BZ}, 
 for the purposes of this Letter, we will not consider it further.
\bibitem{fn2} Models where the $U_{ij}$ are subject to further
  constraints may behave qualitatively differently and are
  not encompassed by the present analysis (e.g. the time-reversal
  invariant counterpart of the RF-model, $U_{ij}\in {\rm O}(N)$).
\end{thebibliography}
\end{document}